\documentclass[12pt]{iopart}

\usepackage{booktabs}
\usepackage{multirow}
\expandafter\let\csname equation*\endcsname\relax
\expandafter\let\csname endequation*\endcsname\relax
\usepackage{amsmath}
\usepackage{amssymb}
\usepackage{graphicx,slashbox,bigstrut,color}
\usepackage{verbatim}
\usepackage{etoolbox}
\usepackage{url}

\begin{document}

\title{Self-paced brain-computer interface control of ambulation in a virtual reality environment}
\author{Po T Wang$^1$, Christine E King$^1$, Luis A Chui$^{2,3}$, An H Do$^{2,3}$ and Zoran Nenadic$^{1,4}$}
\address{$^1$ Department of Biomedical Engineering, University of California, Irvine, CA 92697, USA}
\address{$^2$ Department of Neurology, University of California, Irvine, CA 92697, USA}
\address{$^3$ Division of Neurology, Long Beach Veterans Affairs Medical Center, Long Beach, CA 90822, USA}
\address{$^4$ Department of Electrical Engineering and Computer Science, University of California, Irvine, CA 92697, USA}
\ead{and@uci.edu, znenadic@uci.edu}

\begin{abstract}

\noindent\textit{Objective.}
Spinal cord injury (SCI) often leaves affected individuals unable to ambulate. Electroencephalogramme (EEG) based brain-computer interface (BCI) controlled lower extremity prostheses may restore intuitive and able-body-like ambulation after SCI. To test its feasibility, the authors developed and tested a novel EEG-based, data-driven BCI system for intuitive and self-paced control of the ambulation of an avatar within a virtual reality environment (VRE). 

\noindent\textit{Approach.}
Eight able-bodied subjects and one with SCI underwent the following 10-min training session: subjects alternated between idling and walking kinaesthetic motor imageries (KMI) while their EEG were recorded and analysed to generate subject-specific decoding models. Subjects then performed a goal-oriented online task, repeated over 5 sessions, in which they utilised the KMI to control the linear ambulation of an avatar and make 10 sequential stops at designated points within the VRE. 

\noindent\textit{Main results.}
The average offline training performance across subjects was 77.2$\pm$9.5\%, ranging from 64.3\% (p = 0.00176) to 94.5\% (p = 6.26$\times$10$^{-23}$), with chance performance being 50\%. The average online performance was 8.4$\pm$1.0 (out of 10) successful stops and 303$\pm$53 sec completion time (perfect = 211 sec). All subjects achieved performances significantly different than those of random walk (p$<$0.05) in 44 of the 45 online sessions.

\noindent\textit{Significance.}
By using a data-driven machine learning approach to decode users' KMI, this BCI-VRE system enabled intuitive and purposeful self-paced control of ambulation after only a 10-minute training. The ability to achieve such BCI control with minimal training indicates that the implementation of future BCI-lower extremity prosthesis systems may be feasible.
\end{abstract}

\section{Introduction}
Neurological conditions such as spinal cord injury (SCI) may leave the affected individuals with paraparesis or paraplegia that renders them unable to ambulate. Currently, there are no methods to restore lower extremity motor functions in this population, which inspired the pursuit of alternative substitutive technologies such as robotic exoskeletons~\cite{argo:10}, functional electrical stimulation (FES) systems~\cite{dgraupe:98}, or spinal cord stimulators~\cite{sharkema:11}. A major limitation of these approaches is the absence of direct supraspinal control, which precludes 
them from achieving the much sought-after able-body function. In addition, issues such as manual control, high cost, and unwieldiness may have prevented their widespread adoption.  Consequently, wheelchairs remain the primary means of mobility after SCI. Unfortunately, the absence of lower extremity utilisation associated with wheelchair use causes a wide variety of comorbidities that contribute to the majority of SCI-related medical care costs~\cite{jkschmitt:03, ssabharwal:03, yzehnder:04, mmpriebe:03}. Therefore, to address current shortcomings in the treatment of paraparesis/paraplegia due to SCI, novel brain-controlled prostheses are being actively pursued~\cite{ptwang:10}. 

The integration of brain-computer interfaces (BCIs) with lower extremity prostheses, such as FES, to restore or improve gait function in this population may constitute one such novel approach. At the time of this study, no integrated BCI-lower extremity prosthesis system for independent overground walking has been reported on. Successful implementation of such a system may potentially reduce disability in subjects with SCI, promote their independence and social integration, and reduce the incidence of associated medical comorbidities.  

An ideal 
BCI-lower extremity prosthesis system is envisioned to have intuitive and robust control, as well as minimal user training. For example, an intuitive strategy for control of a BCI prosthesis may include attempted walking or kinaesthetic motor imagery (KMI) of walking. The feasibility of such a system is contingent upon the ability to robustly decode neurophysiological patterns underlying these control strategies in the face of potential cortical reorganisation following SCI. More specifically, functional magnetic resonance imaging (fMRI) studies suggest that brain areas normally associated with motor imagery of the lower extremity movements or gait may diminish, disappear, or shift following paraplegia due to SCI~\cite{Cramer2005, Sabbah2002, Alkadhi2005, Hotz-Boendermaker2008}. This requires that such a BCI system accommodates for each user's potentially unique physiology. In addition, a BCI system must be designed to facilitate rapid user training, thereby promoting widespread adoption of this technology. The authors hypothesise that a data-driven method for extracting subject-specific electrophysiological correlates underlying intuitive BCI control strategies will satisfy the above criteria and facilitate a BCI system that is intuitive, robust, and permits rapid user training.

This article presents a novel electroencephalogramme (EEG)-based BCI system for intuitive, self-paced control of the ambulation of an avatar within a virtual reality environment (VRE). This BCI-controlled walking simulator employs a data-driven, subject-specific EEG decoding model, which enabled 9 subjects (one with paraplegia due to SCI) to use walking KMI to achieve intuitive control of the avatar's ambulation after a very brief training session. This simulator provides a similar, albeit virtual, experience to the operation of a potential BCI-controlled lower extremity prosthesis, without the associated physical risks~\cite{rleeb:07}. In addition, the use of VRE in the context of BCI has been shown to reduce the decoding error~\cite{gpfurtscheller:06a}.     
The ability to rapidly achieve purposeful control of an avatar within the VRE 
represents a necessary step towards successful integration of EEG-based BCI systems and physical prostheses. It also implies the feasibility of envisioned BCI-lower extremity prosthesis systems. Finally, it may in the future act as the first step in training SCI users to operate such prosthesis systems once they become available.

\section{Methods}

\subsection{Overview}
To determine the feasibility of future BCI-lower extremity prosthesis systems for ambulation, a data-driven, subject-specific decoding methodology that enabled intuitive BCI control was utilised. To this end, 8 able-bodied subjects and a single subject with paraplegia due to SCI used walking KMI to operate the ambulation of the avatar within a VRE. They first underwent alternating epochs of walking KMI and idling while their EEG data were collected. Subsequently, a computer algorithm used this training data to extract salient EEG signal features and train an EEG classifier. The training procedure was followed by an online BCI evaluation, where subjects utilised walking KMI and idling to asynchronously control the linear ambulation of an avatar within the VRE. To assess the attainment of purposeful control, subjects' performances were recorded over several online sessions and compared to random walk Monte Carlo simulations. 

\subsection{Subject Recruitment}
The study was approved by the University of California, Irvine Institutional Review Board. Nine subjects were recruited and gave their informed consent to participate. Their demographic data are shown in Table~\ref{tab:ListOfSubjects}.  

\begin{table}[ht]
\caption{List of participants with demographic data and prior BCI experience relevant to the task. SCI status scored according to American Spinal Injury Association (ASIA) Impairment Scale.}
\centering
\begin{tabular}{ccccc}
\br
Subject & Gender & Age & BCI experience & SCI status \\
\mr
A1 & M & 40 & $\sim$1 hr & -- \\
A2 & M & 29 &  $\sim$1 hr & --\\
A3 & F & 23 &  $\sim$1 hr & --\\
A4 & F & 57 &  ~~0 hr & --\\
A5 & F & 24 &  ~~0 hr & -- \\
A6 & M & 21 &  ~~0 hr & -- \\
A7 & M & 25 &  ~~0 hr & -- \\
A8 & M & 32 &  ~~0 hr & -- \\
S1 & F & 27 &  ~~0 hr & T8 ASIA B, 11 yr post injury \\
\br
\end{tabular}
\label{tab:ListOfSubjects}
\end{table}

\subsection{Data Acquisition}

Each subject was seated in a chair approximately one metre from a computer monitor that displayed either textual cues (during training sessions) or the VRE (during online sessions). EEG was recorded using a 63-channel EEG cap (Medi Factory, Heerlen, The Netherlands) with Ag-AgCl electrodes arranged according to the extended 10-20 International Standard. Conductive gel (Compumedics USA, Charlotte, NC) was applied to all electrodes and the 30-Hz impedances between each electrode and the reference electrode were maintained at $<$10~K$\Omega$ by abrading the scalp with a blunt needle.  Two NeXus-32 EEG systems (MindMedia, Roermond-Herten, The Netherlands) were linked together and used to amplify and digitise (sampling rate: 256 Hz, resolution: 22 bits, built-in anti-aliasing filter: 27\% of sampling rate) the EEG signals. Signals were streamed in real-time to a computer and subsequently re-referenced in a common average mode. Data acquisition and analysis were performed using custom-made MATLAB\texttrademark ~(MathWorks, Natick, MA) programmes.

\subsection{Training Procedure}\label{sec:tp}
To facilitate intuitive control of the BCI, subject-specific EEG decoding models were generated to differentiate between EEG underlying idling and walking KMI. In addition to being more intuitive than visual motor imagery, KMI is known to provide better separability of EEG for BCI applications~\cite{cneuper:05}.  To this end, subjects were instructed by textual cues to generate walking KMI (i.e. imagine themselves walking) and idling KMI (i.e. relax), while their EEG data were recorded. The textual cues alternated every 30 sec for a total of 10 min. At the same time, the EEG data were labelled as either walking or idling by a corresponding computer signal recorded by an auxiliary data acquisition system (MP150, Biopac Systems, Goleta, CA). The labelling and EEG signals were synchronised by sending a common pulse train to both the MP150 and $\text{NeXus-32}$ data acquisition systems. Electromyogramme (EMG) activity was not recorded to monitor for minor limb movements, since increased EMGs are often observed during KMI~\cite{Jacobson1930,Wehner1984,Lebon2008,Dickstein2005}. 
Instead, the subjects were instructed to refrain from moving during the training procedure, which was enforced by observing the procedure and discarding the entire session if it was considered contaminated by movements.

\subsection{Offline Signal Analysis and Decoding Model Generation}\label{sub:osaapmg}
The training EEG data were analysed offline to generate a subject-specific decoding model. First, the EEG and labelling signals were aligned using the common synchronisation pulse train. In addition, EEG channels with excessive EMG activity were excluded from further analysis using an iterative artefact rejection algorithm~\cite{ahdo:11}. The pre-processed continuous 10-min EEG record was then split into 30-sec long segments of idling and walking states based on the labelling signal. Due to uncertainties in timing between the computer cue and the subject's reaction, the first 8 sec of each state were removed from analysis. Each remaining 22 sec EEG segment was then divided into five 4-sec long non-overlapping trials for a total of 100 trials. 

The labelled EEG trials were then Fast Fourier Transformed (FFT), and their power spectral densities were integrated in 2-Hz bins that were centred at $1,\,3,\, \cdots,\,39$ Hz, yielding 20 power spectral values per channel. Note that this resulted in high-dimensional data ($\sim$1000 dimensions), which significantly exceeded the number of trials, thereby causing a small sample size problem~\cite{kfukunaga:90}. Therefore, the dimension of the input data was initially reduced using classwise principal component analysis (CPCA)~\cite{Das2007, kdas:09}. The class differences were enhanced and the dimension was further reduced by  
either approximate information discriminant analysis (AIDA)~\cite{znenadic:07,Das2008} or Fisher's linear discriminant analysis (LDA)~\cite{Fisher1936}. The choice between AIDA and LDA was determined based on the average classification accuracy described at the end of this paragraph. The combination of these methods yields a piecewise linear feature extraction mapping that approximately maximises the mutual information between the features and class labels~\cite{znenadic:07}. A detailed account of these techniques can be found in~\cite{Das2007,kdas:09,Das2008}. This resulted in the extraction of one-dimensional (1D) spatio-spectral features: 

\begin{equation}\label{eq:features}
f = \mathbf{T}\mathbf{\Phi}_{\text{C}}(\mathbf{d})
\end{equation}  
where $\mathbf{d}\in \mathbb{R}^{B\times C}$ is a single-trial of EEG data, $B$ is the number of frequency bins per channel, $C$ is the number of retained EEG channels,  $\mathbf{\Phi}_{\text{C}}:\mathbb{R}^{B\times C}\rightarrow \mathbb{R}^{m}$ is a piecewise linear mapping from the data space to an $m$-dimensional CPCA-subspace, and $\mathbf{T}:\mathbb{R}^{m}\rightarrow \mathbb{R}$ is an AIDA or LDA transformation matrix. Once 1D spatio-spectral features were extracted, a linear Bayesian classifier: 

\begin{equation}
\label{eq:lrt}
f^{\star} \in
\begin{cases}
\mathcal{I}, &\text{if}\quad P(\mathcal{I}\,|f^{\star})>P(\mathcal{W}\,|f^{\star})\\
\mathcal{W}, &\text{if}\quad P(\mathcal{W}\,|f^{\star})>P(\mathcal{I}\,|f^{\star})
\end{cases}
\end{equation}
was designed, where $P(\mathcal{I}\,|f^{\star})$ and $P(\mathcal{W}\,|f^{\star})$ are the posterior probabilities of idling and walking classes given the observed feature, $f^{\star}$, respectively, and were calculated using the Bayes rule. Note that the classifier~(\ref{eq:lrt}) utilises the maximum posterior probability (MAP) rule. The classification accuracy of the Bayesian classifier~(\ref{eq:lrt}) was then assessed by performing 10 runs of stratified 10-fold cross-validation (CV)~\cite{rkohavi:95}. 

This above procedure was systematically repeated to find the optimal frequency range~\cite{ahdo:11}. Briefly, the lower frequency bound was increased in 2-Hz steps until the classifier performance stopped improving, allowing the optimal lower frequency bound, $F_L$, to be determined.  The optimal higher frequency bound, $F_H$, was found in a similar manner. The optimal frequency range, the list of retained channels after artefact rejection, the feature extraction mapping, and the classifier parameters---referred to as the decoding model, were then saved for real-time EEG analysis. Finally, the signal processing, feature extraction, and classification algorithms were implemented into the BCI software and optimised for real-time operation.

\subsection{Online Signal Analysis}\label{sub:osa}
During online operation, blocks of EEG data were acquired every 0.5 sec. This rate was limited by the computer processing speed and was empirically found to ensure data acquisition without dropping  packets. The EEG data were then divided into 0.75-sec long segments and were processed as described in Section~\ref{sub:osaapmg}. Note that this segment length provided 
an accurate estimation of EEG spectral power even at the lower end of physiologically relevant frequencies~\cite{ahdo:11}. Subsequently, the EEG signals from artefact-prone channels were excluded, and the remaining EEG data were transformed into the frequency domain by FFT. The power spectral densities over the optimal frequency range were calculated and used as an input to the feature extraction algorithm~(\ref{eq:features}). The posterior probabilities of idling and walking classes given the observed EEG feature were then calculated using the Bayes rule.

\subsection{Online Calibration}\label{sec:oc}
Prior to online BCI operation, a short calibration procedure was performed to determine state transition rules suitable for self-paced online BCI operation. This is necessary because unlike offline analysis that is based on well-segmented and labelled EEG trials, online data segments may lie at class transitions. This would cause the MAP rule~(\ref{eq:lrt}) to create an excessively noisy state transition sequence, which may frustrate the user during online BCI operation. A similar calibration approach was found to be effective in related self-paced BCI studies~\cite{ptwang:10, ahdo:11,ceking:11}. 

\begin{figure}[!htbp]%
\includegraphics[width=\columnwidth]{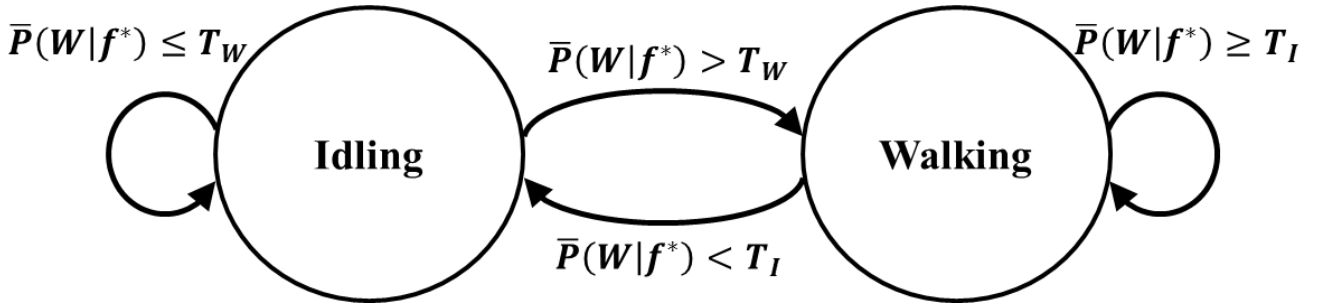}%
\caption{The BCI system is a binary state machine with idling and walking KMI states represented by circles. The state transitions are represented by arrows, with transitions triggered by the conditions shown next to the arrows. Self-pointing arrows denote that the system remains in the present state.}%
\label{fig:statemachine}%
\end{figure}

The self-paced BCI operation is modelled as a binary state machine (see Figure~\ref{fig:statemachine}), where state transitions are triggered by comparing the posterior probabilities to suitably chosen thresholds, $T_{\mathcal{I}}$ and $T_{\mathcal{W}}$. The system transitions from the idling to walking state when $\bar{P}(\mathcal{W}\,|f^{\star})> T_{\mathcal{W}}$, where $\bar{P}(\mathcal{W}\,|f^{\star})$ is the posterior probability of the walking class given the observed feature, $f^{\star}$, averaged over the 
most recent 1.5 sec of EEG data (note that averaging may further smooth the state transitions). Conversely, the system transitions from the walking to idling state whenever $\bar{P}(\mathcal{W}\,|f^{\star})< T_{\mathcal{I}}$. When $T_{\mathcal{I}}\le\bar{P}(\mathcal{W}\,|f^{\star})\le T_{\mathcal{W}}$, the system remains in the present state. In summary, unlike the MAP rule~(\ref{eq:lrt}) that essentially uses $T_{\mathcal{I}}=T_{\mathcal{W}}=0.5$, the proposed scheme requires more substantial evidence before state transitions are initiated. If not, the default behaviour of the system is to remain in the present state, which also reduces the subject's mental workload. 

To determine the optimal thresholds, the BCI system ran in the online mode while subjects were prompted to alternate between idling and walking KMI for a total of $\sim$2 min. During each mental state, the posterior probabilities were calculated as in Section~\ref{sub:osa}, and their histogrammes were plotted. Based on these histogrammes, the thresholds were initially chosen as: $T_{\mathcal{W}} = \text{median}\left\{\bar{P}(\mathcal{W}\,|f^{\star}\in \mathcal{W})\right\}$, and $T_{\mathcal{I}} = \text{median}\left\{\bar{P}(\mathcal{W}\,|f^{\star}\in \mathcal{I})\right\}$, where $\bar{P}(\mathcal{W}\,|f^{\star}\in \mathcal{W})$ and $\bar{P}(\mathcal{W}\,|f^{\star}\in \mathcal{I})$ represent the posterior probabilities of walking given that the subject was instructed to engage in walking KMI and idling, respectively. A short online test was then performed and based on the subject's feedback, these initial threshold values were further adjusted prior to online operation in order to help optimise the performance.

While this approach may appear similar to the concept of a ``brain switch''~\cite{gebirch:02,grpfurtscheller:10a}, there are significant differences between the two approaches. For example, the brain switch toggles between idle and active states with a single motor imagery, and once the system switches from one state to another, the motor imagery is no longer required. On the other hand, our system requires two motor imageries, and to remain in the present state, the corresponding motor imagery must be sustained. While this may increase the mental workload of the subject, this approach is more intuitive because of the 1:1 correspondence between motor imageries and intended actions. 

\subsection{BCI and VRE Integration}
\begin{figure}%
\includegraphics[width=\columnwidth]{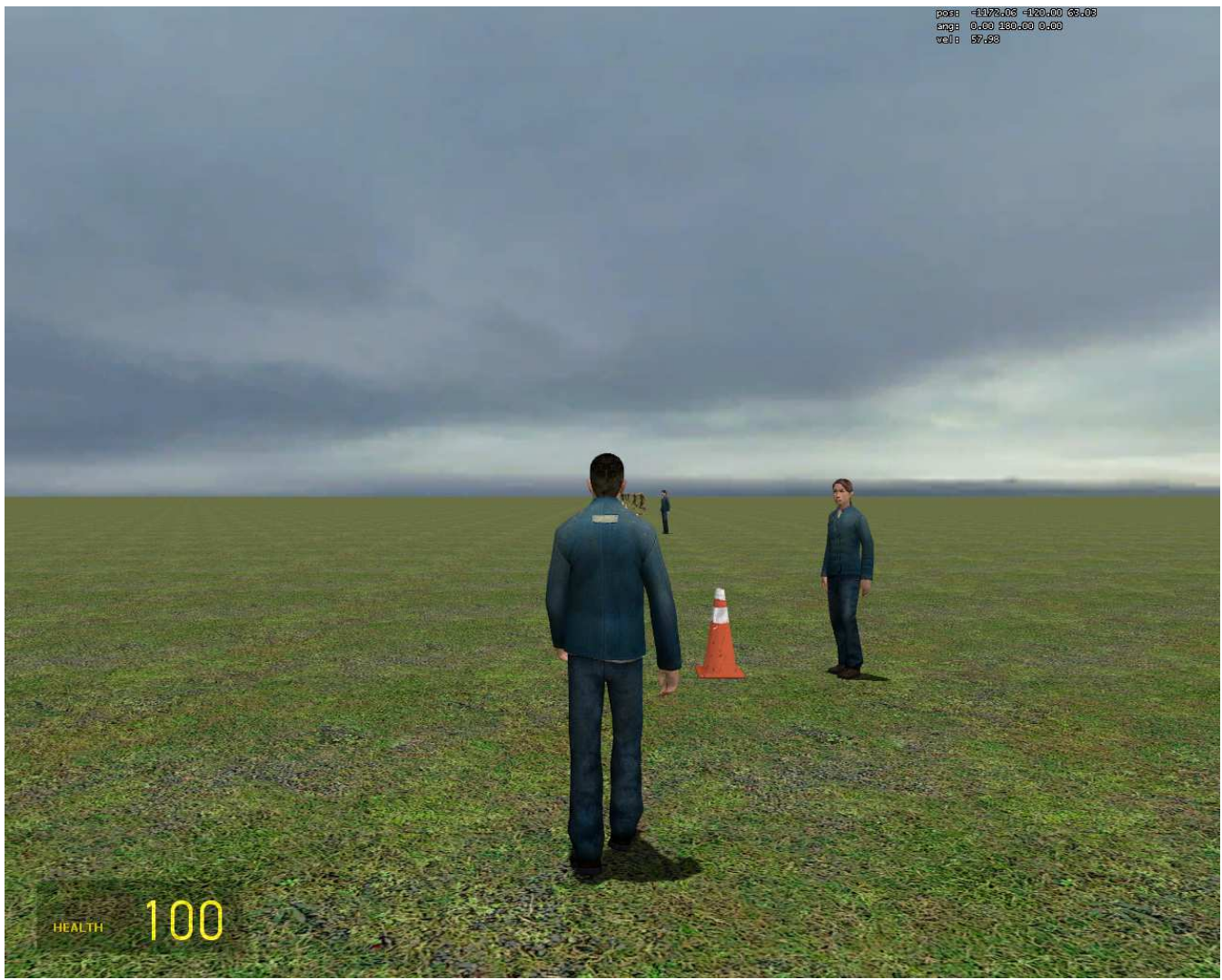}%
\caption{The VRE with the BCI-controlled avatar in 3rd person ``over-the-shoulder'' view. Shown next to the avatar is an NPC and a traffic cone. The position/speed readouts are shown in the top right corner.}%
\label{fig:garry_screenshot}%
\end{figure}

The VRE was constructed using Garry's Mod\texttrademark ~simulated physics environment (Valve Corporation, Bellevue, WA), and consisted of a flat grassland with 10 non-player characters (NPCs) standing in a straight line. The course length was $\sim$120 body lengths ($\sim$210 m, assuming a body length of 1.75 m) along the users' avatar linear path (see Figure~\ref{fig:garry_screenshot}). 
This design is intended to facilitate a goal-oriented online test in which the subjects utilised walking KMI and idling to walk the avatar forward and stop by each NPC, similar to Leeb et al.~\cite{rleeb:07}. Further details of the online evaluation are described in Section~\ref{sec:opaa}.

To interface the BCI software and VRE,  a virtual joystick programme (Parallel Port Joystick~\cite{DvanderWesthuysen:2011}) was used. To this end, a C$++$ dynamic-link library was developed to relay BCI commands to  move/stop the avatar via the virtual joystick. Finally, a custom-made C\# programme  performed optical character recognition on the position readouts from the VRE's display (see Figure~\ref{fig:garry_screenshot}) in order to automatically track the subject's online BCI performance.

\subsection{Online Performance and Assessment}\label{sec:opaa}
To assess the online BCI performance, subjects used walking KMI to move the avatar to each NPC and idling KMI to stand still within a two-body length radius (centred at NPC) for at least 2 sec. A short video demonstrating the task can be found at \htmladdnormallink{http://youtu.be/GXmovT3BxEo}{http://youtu.be/GXmovT3BxEo}. Similar to the training procedure, the subjects were instructed to refrain from moving and were asked to repeat the task if movements were detected. Each subject repeated this task over 5 sessions, with all sessions completed within a single day. Two performance measures were recorded during each session: the time taken to complete the course and the number of successful stops. Subjects received one point for idling the avatar within the designated stop for at least 2 sec; therefore, the maximum successful stop score was 10 points. In addition, only a fraction of the point was awarded for dwelling between 0.5 and 2.0 sec. Note that subjects were not penalised for dwelling longer than 2 sec, however, this will inevitably increase the completion time and therefore lower the overall performance. A 20-min time limit was enforced, beyond which the online session was interrupted and the number of successful stops achieved thus far was recorded. Ideally, it should take on average 18 sec to walk from one NPC to the next without stopping, with the total course completion time of 211 sec (191 sec for walking and 20 sec for idling).  

\subsubsection{Control Experiments:}\label{sec:random}
The numbers of successful stops and completion times were compared to those achieved by random walk to determine whether purposeful control was attained. Random walk performances were simulated by sampling the posterior probabilities uniformly between 0 and 1 and applying the state transition rules outlined in Section~\ref{sec:oc} with the subject-specific threshold values, $T_{\mathcal{I}}$ and $T_{\mathcal{W}}$. The random walk simulator was also allotted the 20-min time limit, and the number of successful stops was calculated in the same manner as above. To facilitate statistical testing, 1000 Monte Carlo runs of the random walk simulation were performed. The subjects' performances were then compared to those of the Monte Carlo simulation, and empirical p-values were calculated. An additional control experiment consisted of an able-bodied subject manually performing the same task with a physical joystick. 

\subsubsection{Statistical Tests:}\label{sec:pvalue}
The 2D probability density function (PDF), with number of successful stops and completion times as variables, of each subject's simulated random walk was estimated using the Parzen window method~\cite{roduda:01,botev2006}.  Through each subject's observed performance point (consisting of a successful stops and completion time pair), a constant-value PDF contour was drawn. The volume under the PDF outside the contour was then found by numerical integration, effectively defining the p-value (the null hypothesis being that the subjects' performances are no different from random walk). Purposeful control was defined as the ability to complete the task within 20 min with performances significantly different from random walk in a multivariate analysis.

\section{Results}\label{sec:results}

\subsection{Offline Performance}\label{sec:op}
The 9 subjects underwent training data collection as described in Section~\ref{sec:tp}, and subject-specific EEG decoding models were generated as described in Section~\ref{sub:osaapmg}. Cross-validation of these models resulted in classification accuracies ranging from 64.3\% to 94.5\% (Table~\ref{tab:OfflinePerformances}), with p-values $<$ $0.01$ (the null hypothesis being defined as having a chance level classification accuracy of 50\%). The average offline performance of the able-bodied subjects was 75.1\%, compared to the 94.5\% accuracy of the SCI subject.

\begin{figure}%
\includegraphics[width=\columnwidth]{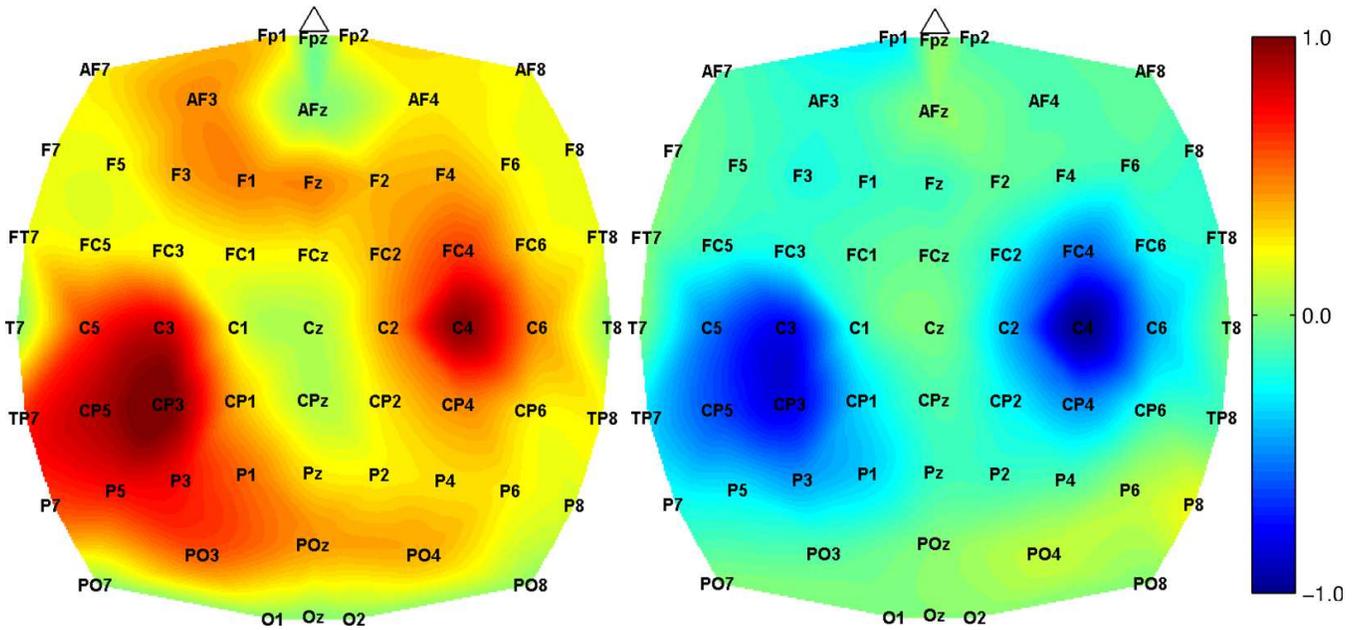}%
\caption{Spatio-spectral feature extraction maps corresponding to the 12--14 Hz frequency range for Subject~A2. Dark colours (red and blue) represent the areas that were responsible for encoding the differences between idling and walking KMI. Since the feature extraction method is piecewise linear (see Section~\ref{sub:osaapmg}), there are two maps: the map on the left (right) corresponds to the subspace adapted to the idle (walking) class, respectively.}%
\label{fig:feimags01}%
\end{figure}

\begin{figure}%
\includegraphics[width=\columnwidth]{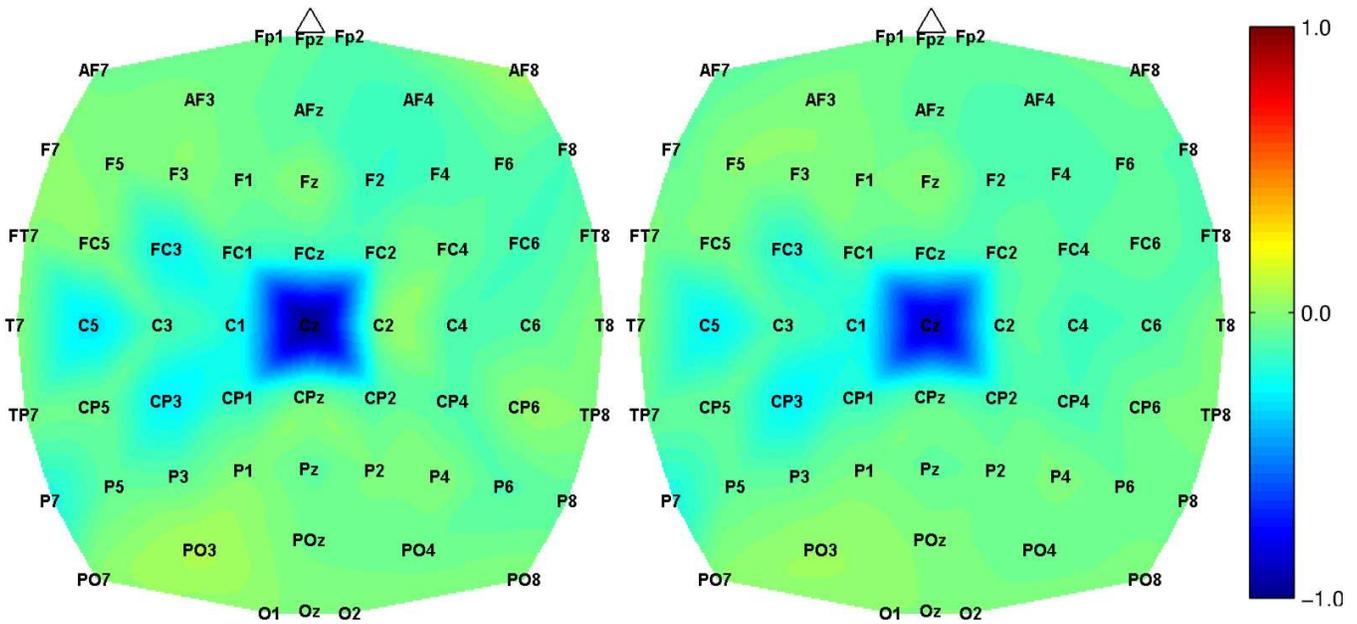}%
\caption{Feature extraction maps of Subject~S1 (SCI subject) showing that EEG power in the 14-16 Hz bin at channel Cz was the most informative for encoding the differences between idling and walking classes.}%
\label{fig:feimags04}%
\end{figure}

\begin{table}[ht]
\caption{Offline performances represented as classification accuracies estimated with 10 runs of stratified 10-fold CV. The classification accuracy is defined as the probability of correctly classifying a trial given the feature, $f^{\star}$, i.e. $P(\text{correct}\,|f^{\star})=P(\mathcal{I}\,|f^{\star}\in\mathcal{I})P(\mathcal{I})+ P(\mathcal{W}\,|f^{\star}\in\mathcal{W})P(\mathcal{W})$, where $P(\mathcal{I}\,|f^{\star}\in\mathcal{I})$ and $P(\mathcal{W}\,|f^{\star}\in\mathcal{W})$ are defined in Section~\ref{sub:osaapmg}, and $P(\mathcal{I})$ and $P(\mathcal{W})$ are the prior probabilities of idling and walking classes, respectively. The number of retained channels (RC) after artefact rejection and the optimal frequency range corresponding to each subject's offline performance are also included. $^{\dagger}$32-channel EEG montage was used due to technical difficulties.}\label{tab:OfflinePerformances}
\centering
\begin{tabular}{crllc}
\br
Subject & $P(\text{correct}\,|f^{\star})$ & p-value & RC & Freq. band \\
\mr
A1 & 88.3 $\pm$ 0.7\% & $ 1.27 \times 10^{-16} $ & 54 & 6 -- 20 Hz \\ 
A2 & 86.6 $\pm$ 0.8\% & $ 6.56 \times 10^{-15} $ & 54 & 8 -- 24 Hz \\ 
A3 & 76.0 $\pm$ 1.3\% & $ 9.05 \times 10^{-8} $  & 54 & 6 -- 20 Hz \\ 
A4 & 80.9 $\pm$ 1.2\% & $ 1.35 \times 10^{-10} $ & 32$^{\dagger}$ & 4 -- 40 Hz \\ 
A5 & 67.4 $\pm$ 2.2\% & $ 2.04 \times 10^{-4} $  & 54 & 8 -- 40 Hz \\ 
A6 & 72.5 $\pm$ 1.6\% & $ 2.35 \times 10^{-6} $  & 42 & 4 -- 18 Hz \\ 
A7 & 64.3 $\pm$ 1.1\% & $ 1.76 \times 10^{-3} $  & 50 & 6 -- 40 Hz \\ 
A8 & 64.5 $\pm$ 1.8\% & $ 1.80 \times 10^{-3} $  & 25$^{\dagger}$ & 4 -- 40 Hz \\ 
S1 & 94.5 $\pm$ 0.8\% & $ 6.26 \times 10^{-23} $ & 53 & 8 -- 40 Hz \\ 
\mr
A1-8 &  75.1 $\pm$ 9.5\% & ~~~-- & --- & ---\\
All  & 77.2 $\pm$ 11.0\% & ~~~-- & --- & ---\\
\br
\end{tabular}
\end{table}
Further analysis of the subject-specific feature extraction maps demonstrated that the most informative features for classification in able-bodied subjects were the EEG powers in the 4--18 Hz frequency range over the lateral central/centro-parietal areas (see Figure~\ref{fig:feimags01}). However, for the SCI subject (Subject~S1), 
the EEG powers in the 14--18 Hz frequency range over the mid-central areas were the most informative features for classification (see Figure~\ref{fig:feimags04}). 

\subsection{Online Calibration}\label{sec:oc1}
\begin{figure}[h]
        \centering
                \includegraphics[width=0.85\columnwidth]{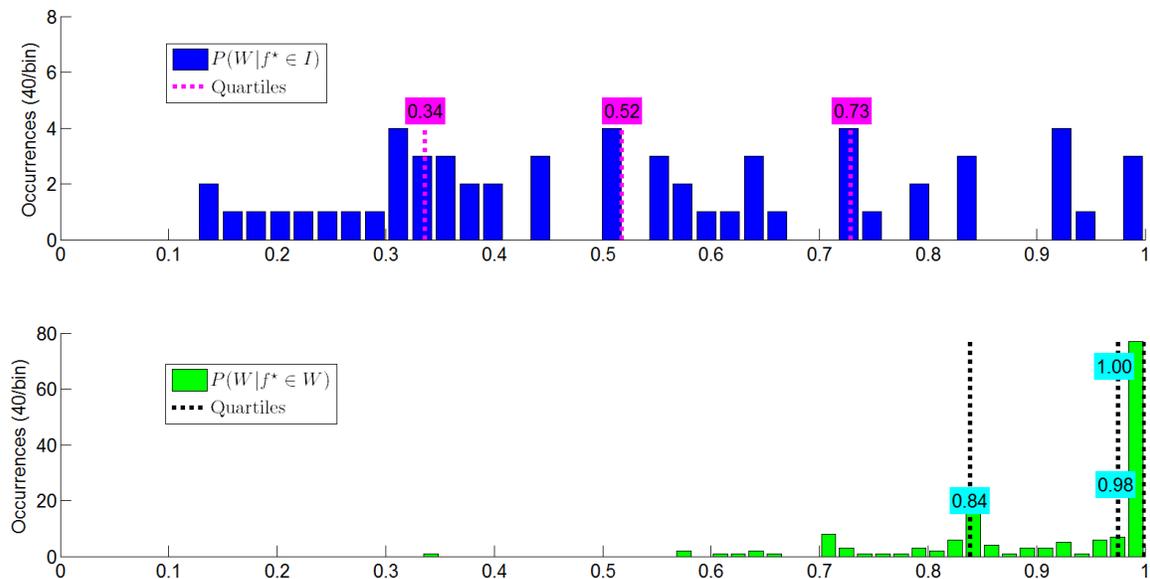}
        \caption{Histogramme of the distribution of posterior probabilities of idling (top) and walking KMI (bottom) states for Subject~S1. The dotted lines denote quartiles.}
        \label{fig:histogram}
\end{figure}

After a short calibration procedure (described in Section~\ref{sec:oc}), the distributions of the posterior probabilities, $P(\mathcal{W}|f^{\star})$,  were estimated as histogrammes (see Figure~\ref{fig:histogram}). Note that in the ideal situation, the distribution of $P(\mathcal{W}|f^{\star}\in\mathcal{W})$ should cluster around 1, whereas $P(\mathcal{W}|f^{\star}\in\mathcal{I})$ should cluster around 0. 
In reality, due to the inherent noise in EEG, these distributions will have some overlap. 
The state transition thresholds, $T_{\mathcal{I}}$ and $T_{\mathcal{W}}$, were then determined and their values are shown in Table~\ref{tab:thresholds}. Also, if EEG data carried no class information, the two thresholds would be equal to $P(\mathcal{W})$ (0.5 in the present study). On the other hand, if classes could be perfectly decoded from EEG data, the threshold values $T_{\mathcal{I}}$ and $T_{\mathcal{W}}$ would approach 0 and 1, respectively. The values of $T_{\mathcal{I}}$ and $T_{\mathcal{W}}$ varied  across subjects from 0.19 to 0.55 and from 0.43 to 0.91, respectively, and as apparent in Table~\ref{tab:thresholds}, the thresholds for all subjects were separable. In addition, the calculated values of $T_{\mathcal{W}}$ were found to correlate with the offline performances shown in Table~\ref{tab:OfflinePerformances} ($\rho$ $=$ 0.87, p $=$ 0.002). However, the same was not true for   $T_{\mathcal{I}}$ ($\rho$ $=$ 0.05,  p $=$ 0.90). Finally, it was found that the offline performances also correlate with the separability of $T_{\mathcal{W}}$ and $T_{\mathcal{I}}$ (i.e. $T_{\mathcal{W}}$ $-$ $T_{\mathcal{I}}$), resulting in a correlation coefficient of 0.80 (p $=$ 0.009). Note, however, that these tests were based on only~9 samples. 
\begin{table}[!htbp]
\caption{The chosen values of thresholds $T_{\mathcal{I}}$ and $T_{\mathcal{W}}$.}
\centering
\begin{tabular}{lcc}
\br
Subject &  $T_{\mathcal{I}}$  & $T_{\mathcal{W}}$ \\ \mr
A1 & 0.53 & 0.91\\ 
A2 & 0.24 & 0.64\\ 
A3 & 0.19 & 0.56\\ 
A4 & 0.43 & 0.58\\ 
A5 & 0.55 & 0.57\\ 
A6 & 0.53 & 0.61\\ 
A7 & 0.41 & 0.43\\
A8 & 0.19 & 0.45\\
S1 & 0.32 & 0.87\\ 
\br
\end{tabular}
\label{tab:thresholds}
\end{table}

\subsection{Online Performance}\label{sec:online}
\begin{figure}[!htbp]
        \centering
                \includegraphics[width=1.00\textwidth]{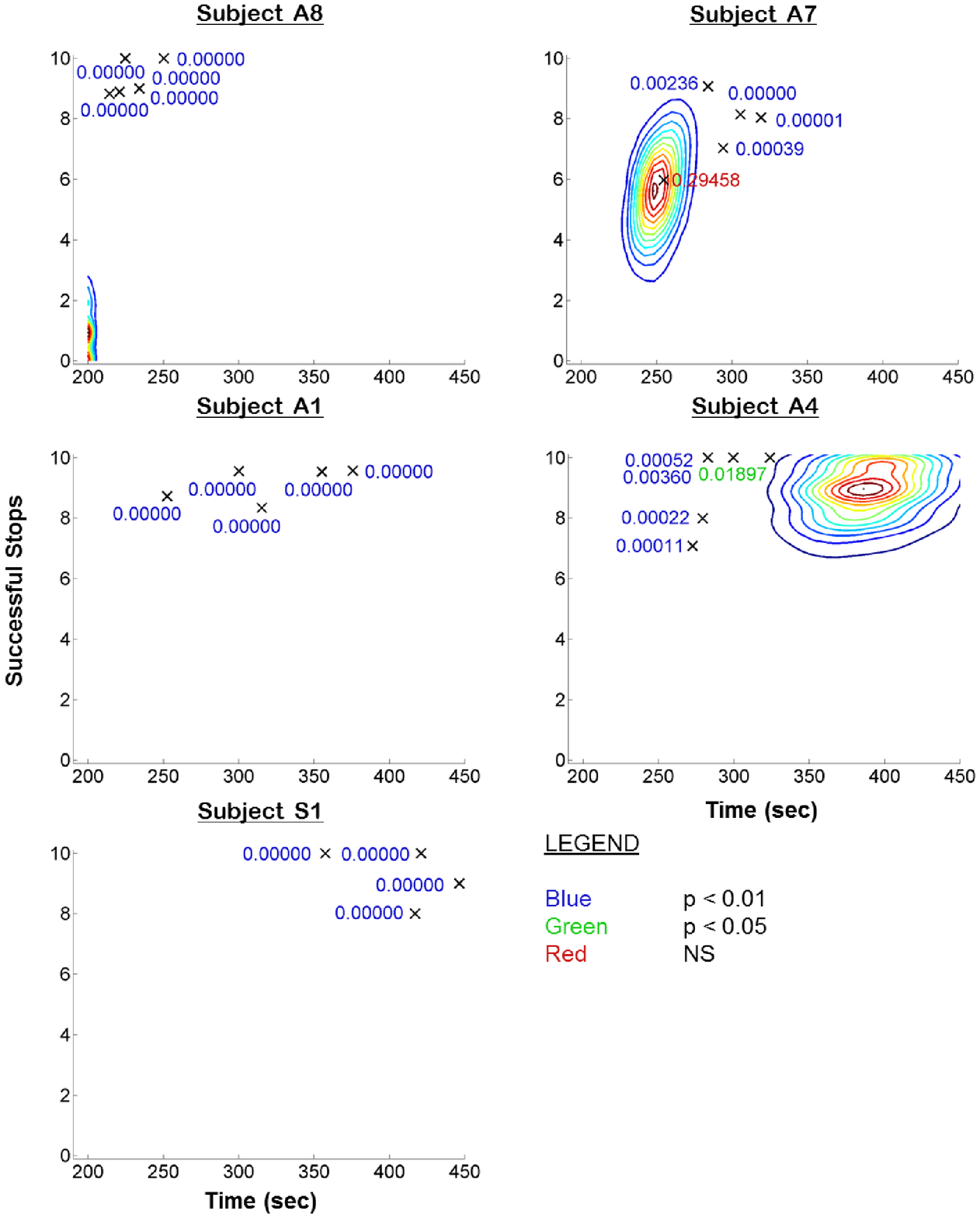}
     \caption{Online performances of 5 representative subjects. Each cross corresponds to one online session's completion time and successful stops. The numbers next to the crosses indicate the empirical p-value, calculated as described in Section~\ref{sec:pvalue}, and are colour coded by their significance level (see legend). The random walk PDFs are displayed as contours. 
        Subjects~A8 (A7) had the best (worst) online performances, respectively. 
 For Subjects~A1 and~S1, the random walk simulation did not complete the task within the allotted 20 minutes, hence contours are not shown.}
        \label{fig:MC_results}
\end{figure}

As described in Section~\ref{sec:opaa}, the online performances of all subjects operating the BCI-controlled walking simulator were evaluated by comparing the task completion times and the successful stops to those of the simulated random walk. After completing 1000 Monte Carlo random walk trials per subject with the threshold parameters given in Table~\ref{tab:thresholds}, the 2D PDF contours were constructed and the empirical p-values of the subject's online performances (numbers of successful stops and completion times) were calculated. The online performances, the corresponding empirical p-values, and the random walk PDF contours of 5 representative subjects (including the SCI subject) are shown in Figure~\ref{fig:MC_results}. Overall, in 43 out of 45 online sessions, subjects achieved performances that were significantly different (i.e. ``outside of the contours'') from those of random walk (p$<$0.01). At a significance level of 0.05, performances were different from random walk in 44 sessions (Subject~A7 had a single session with a non-significant performance). 
In addition, the average completion times and successful stops are summarised in Table~\ref{tab:OnlinePerformances}. Note that the completion time consists of a fixed walking time (191 sec) and a variable amount of idling time.   

\begin{table}[!htbp]
\caption{Average online performances of the subjects compared to those of random walk (denoted by RW in the immediate row).} 
\label{tab:OnlinePerformances}
\centering
\begin{tabular}{lccc}
\toprule
          Subject  &  Completion Time (sec)   &   Successful Stops   \\
                   &  mean~$\pm$~std          &   mean~$\pm$~std     \\
\midrule
          A1  &  319.8 $\pm$ 48.3        &  9.14 $\pm$ 0.57   \\
       A1-RW  &          $> 1200$        &  0.24 $\pm$ 0.43   \\
\midrule
          A2  &  266.7 $\pm$ 10.8        &  7.80 $\pm$ 1.10  \\
       A2-RW  &  224.4 $\pm$ 18.1        &  1.47 $\pm$ 1.09  \\
\midrule
          A3  &  291.9 $\pm$ 19.3        &  8.03 $\pm$ 1.08  \\
       A3-RW  &  219.3 $\pm$ ~9.8        &  2.46 $\pm$ 1.37  \\
\midrule
          A4  &  291.7 $\pm$ 20.6        &  9.01 $\pm$ 1.39  \\
       A4-RW  &  383.4 $\pm$ 26.7        &  9.15 $\pm$ 0.82  \\
\midrule
          A5  &  325.4 $\pm$ 54.2        &  8.10 $\pm$ 0.94  \\
       A5-RW  &  602.6 $\pm$ 38.5        &  9.89 $\pm$ 0.24  \\
\midrule
          A6  &  318.2 $\pm$ 27.3        &  8.09 $\pm$ 1.06  \\
       A6-RW  &  699.0 $\pm$ 53.4        &  9.91 $\pm$ 0.24  \\
\midrule
          A7  &  291.5 $\pm$ 24.4        &  7.65 $\pm$ 1.19  \\
       A7-RW  &  251.6 $\pm$ 9.8         &  5.62 $\pm$ 1.24  \\
\midrule
          A8  &  228.8 $\pm$ 14.0        &  9.34 $\pm$ 0.60  \\
       A8-RW  &  193.9 $\pm$ 2.3         &  0.39 $\pm$ 0.54  \\
\midrule
          S1  &  410.6 $\pm$ 37.4        &  9.25 $\pm$ 0.96  \\
       S1-RW  &          $> 1200$  &  4.52 $\pm$ 1.60  \\
\midrule
          Able-bodied     &  292.4 $\pm$ 41.4   &  8.39 $\pm$ 1.12  \\
          All subjects      &  302.9 $\pm$ 53.0    &  8.46 $\pm$ 1.12  \\
\midrule
         Joystick  &  205.07 $\pm$ 4.2  &  9.38 $\pm$ 0.85  \\
\bottomrule
\end{tabular}
\end{table}

As previously mentioned, based on the definition in Section~\ref{sec:pvalue}, purposeful BCI control of the avatar was achieved by all subjects 
in 44 out of 45 online sessions. The performance breakdown according to individual measures is as follows. Subjects~A1 and~S1 achieved purposeful control with superior performance in both measures. Subjects~A4,~A5 and~A6 achieved purposeful control with superior performance in completion time only. On the other hand, Subjects~A2,~A3, A7 and A8 achieved purposeful control with superior performance in the number of successful stops, although they required more time to complete the task. However, it is crucial that these individual performance measures be interpreted in the context of each other in order to be meaningful, and these points will be further elaborated upon in Section~\ref{sec:discussion}. Finally, to demonstrate the performance level achievable by manual control, an able-bodied subject performed the task with a physical joystick. The manual joystick performance was significantly different and superior to the BCI performances in terms of completion times (p$=$0.0002) but was not different in terms of the number of successful stops (p$=$0.086). 

\begin{figure}[!htbp]
        \centering
                \includegraphics[width=0.50\textwidth]{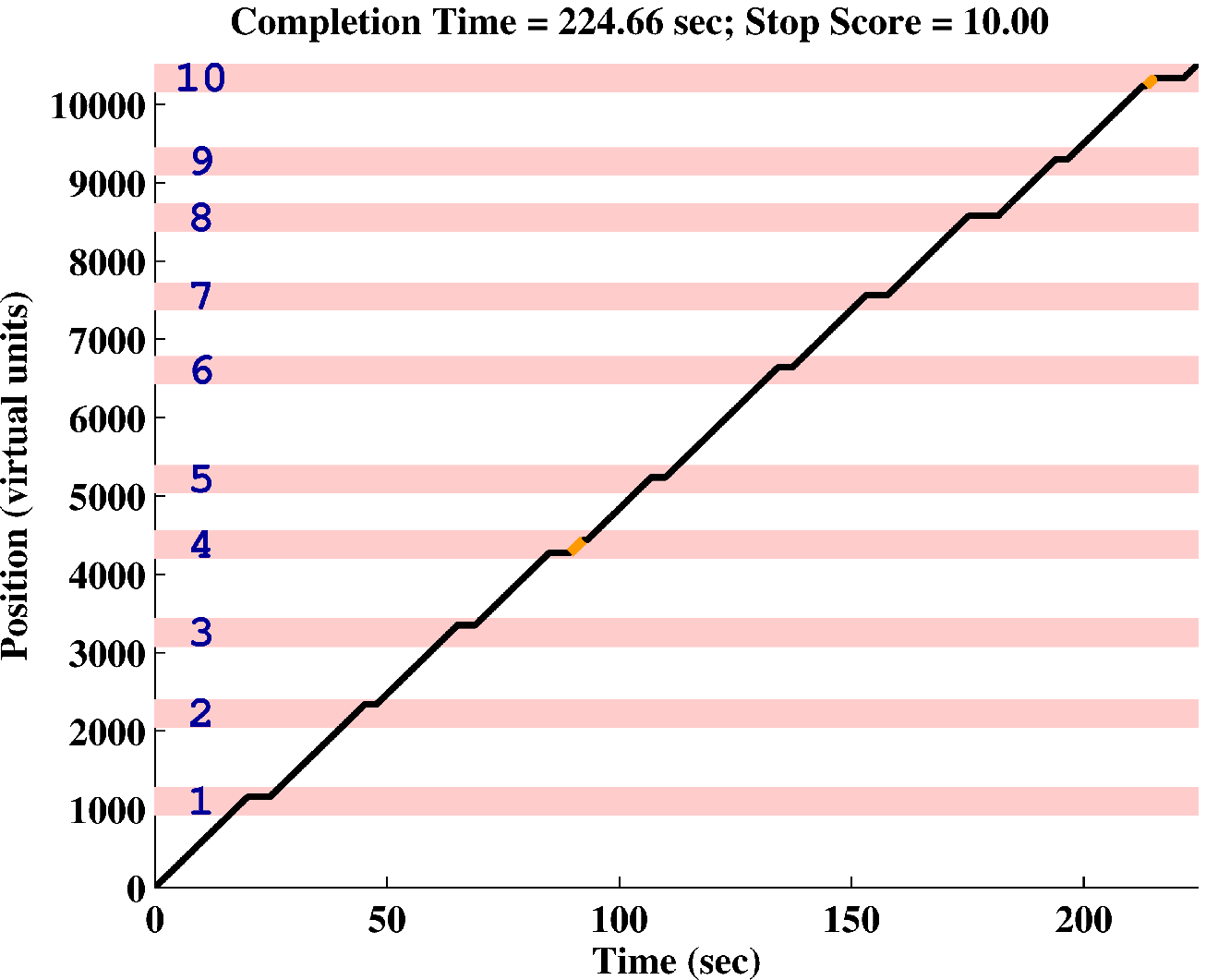}
        \caption{Time-space course of a representative online session for Subject~A8. The pink areas mark designated stopping zones. Orange segments mark false starts. In order to finish the course, a subject is required to walk out of the last stopping zone.}\label{fig:best_77_online}
\end{figure}

Figure~\ref{fig:best_77_online} shows a representative time-space course of one online session for Subject~A8. In this session, not only did the subject complete the course with the maximum successful stops and a short completion time, but also he had only 2 false starts (moving when not supposed to) and no false stops (stopping when not supposed to). Over 5 online sessions, this subject averaged 0.4 false starts and 2.6 false stops. By factoring in the duration of false starts and stops, as well as the completion time, these correspond to error rates of 0.42\% and 3.34\%, respectively. The online performances of the other subjects were not recorded with this level of detail (the computer code was modified in the late stage of the study to accommodate for this function), and so it is not possible to state their exact false start and false stop rates. Since Subject~A8 achieved the best online performance, it is likely that other subjects' online error rates were higher than the numbers reported above. Therefore, while not universally applicable, the results presented in Figure~\ref{fig:best_77_online} illustrate the level of control achievable by this BCI system.      

Finally, multiple linear regression was used to evaluate the correlation between offline and online performances. More specifically, offline performance was treated as a dependent variable with completion times and numbers of successful stops as the regressors. A moderately positive correlation was found with respect to both regressors with a goodness of fit $R^2=0.2091$ (p $=$ 0.0073).

\section{Discussion and Conclusions}\label{sec:discussion}

This study reports on the successful implementation of a self-paced BCI-controlled walking simulator in which 8 able-bodied and 1 SCI subjects acquired intuitive purposeful BCI control of the avatar's linear ambulation after only a 10-min training session followed by a 2-min calibration session. The rapid training and acquisition of purposeful BCI control were facilitated by using a data-driven machine learning method to generate subject-specific EEG decoding models. The decoding models were validated first offline and then during online BCI operation. These results indicate that the system reported here may satisfy the proposed requirements of an ideal BCI-lower extremity prosthesis (robustness, intuitiveness, and short training time) by using a data-driven machine learning method, and it may be feasible to implement such a system in the near future. 

With the exception of the authors' preliminary work~\cite{ptwang:10}, this study represents the first demonstration of integrating an EEG-based BCI with a VR walking simulator. A comparison between this study and related BCI-VRE studies~\cite{gpfurtscheller:06a,rleeb:07} is given in Table~\ref{tab:comparison}. Note that the present approach utilises KMI of walking/idling as a control strategy, which intuitively matches the task. On the other hand, the study in~\cite{rleeb:07}, and especially the one in~\cite{gpfurtscheller:06a}, were less intuitive. Furthermore, the present approach requires significantly shorter training time before the subjects are able to gain online BCI control. With the present approach, both BCI-na\"{i}ve and BCI-experienced subjects were able to achieve purposeful online BCI control within minutes as opposed to months required in the other studies. In addition, this system has been tested in a substantially larger population of subjects, suggesting that it may generalisable. 
Finally, a direct comparison of the results between the present study and related BCI-VRE studies is not possible due to variations in experimental designs.          
\begin{table}[htbp]
        \centering
        \caption{Comparison of the present study to similar studies in the field.}
        \label{tab:comparison}
                \begin{tabular}{llll}\toprule
                        Study & Mental Strategy & Training Time & Sample Size \\ \midrule
                        Present & KMI of walking/idling & 10 min & 8 AB, 1 SCI \\  
                        Wang et al.~\cite{ptwang:10} & KMI of walking/idling & 12 min & 3 AB\\ 
                        Pfurtscheller et al.~\cite{gpfurtscheller:06a} & KMI of foot/hand movement & 3--5 mo & 3 AB\\ 
                        Leeb et al.~\cite{rleeb:07} & KMI of foot movement/idling & 4 mo & 1 SCI\\ \bottomrule
                \end{tabular}
\end{table}

\subsection{Offline Performance}
It follows from Table~\ref{tab:OfflinePerformances}
that idling and walking KMI states could be decoded from underlying EEG signals with moderate to high accuracy. Furthermore, a 10-min training session was sufficient for a data-driven machine learning algorithm to generate subject-specific EEG decoding models. These models achieved offline classification accuracies between 64.3\% and 94.5\% (mean: 77.2\%), which were significantly superior to random chance (50\%) with p-values as small as $10^{-23}$. This was true even for BCI-na\"{\i}ve subjects. Also, note that the subject with SCI achieved the best offline performance, which may be encouraging for the development of future BCI-lower extremity prostheses in this population.

The EEG decoding models also yielded feature extraction maps (e.g. Figs.~\ref{fig:feimags01}~and~\ref{fig:feimags04}) that could be used to uncover the brain areas and frequency bands that differentiate the idling and walking KMI behaviours. In the able-bodied subjects, the EEG features responsible for encoding the differences between the two states were the powers in the $\mu$ and $\beta$ EEG bands from the lateral central and lateral centro-parietal electrodes. The activity measured by these electrodes is most likely localised to the lateral sensorimotor cortex, which is typically associated with hand and arm movements. On the other hand, the most informative features for the SCI subject were the EEG powers in the $\mu$ and $\beta$ bands over the mid-central electrodes. These electrodes are likely to record the activity originating from the medial sensorimotor cortex, where the leg and foot cortical representation areas are classically located. Hence, it appears that there is a divergence between the brain areas employed by able-bodied subjects and the SCI subject while undergoing the walking KMI. It is possible that these differences are simply caused by different mental strategies employed by the subjects. This may be because unlike simple motor imageries often used in BCI studies, such as fist clenching~\cite{djmcfarland:00, gpfurtscheller:06} or foot tapping~\cite{gpfurtscheller:06}, walking KMI emulates a highly complex set of upper and lower extremity movements for which there may not be a universal motor imagery strategy. Based on their feature extraction maps, it can be hypothesised that able-bodied subjects in this study predominantly imagined the arm swinging process of walking as opposed to the leg movement component. It is also possible that due to the extremely small sample size (only 100 training EEG trials), the relative contribution of other potential brain representation areas (e.g. lower extremity motor areas) was masked by a more dominant arm swing imagery in these maps. In contrast, due to complete motor paraplegia in the SCI subject, walking KMI may be a mental task that is as vivid as attempted leg movements or executed walking. These hypotheses could not be formally tested given the limited population size. Finally, given that the important areas of these feature extraction maps are highly localised, it may be possible to further reduce the number of EEG channels from 25--54 (see Table~\ref{tab:OfflinePerformances}) to $\sim$20 or less.    

The proposed data-driven machine learning methodology was able to produce subject-specific decoding models that accommodate for the neurophysiological variations across subjects. This is especially important for BCI users with SCI due to potential post-injury cortical reorganisation.
Namely, recent fMRI studies~\cite{Cramer2005, Sabbah2002, Alkadhi2005, Hotz-Boendermaker2008} report on significant changes in motor cortical representation areas for lower extremity motor imagery following SCI. 
Thus, the involvement of classical walking KMI representation areas seen in Subject~S1 may not be universally present in SCI individuals. This further underscores the importance of a data-driven EEG decoding model. On the other hand, its lack of specificity may mean that the optimal spatio-spectral features identified by this model are not exclusively associated with walking, but also with non-ambulatory leg or foot movements. Hence, additional studies are necessary to better pinpoint the source and nature of neurophysiological signals underlying both walking KMI and attempted walking in this population. Given the limited signal-to-noise ratio and resolution of EEG, this feat may require the use of invasive recording modalities.

\subsection{Online Calibration}
The state transition thresholds determined in the online calibration session (Table~\ref{tab:thresholds}) demonstrated that the transitions from idling to walking states (and vice versa) were highly separable. Despite the limited sample size, the values of $T_{\mathcal{W}}$ appeared to exhibit a positive correlation trend with the offline performances.
This further validates the proposed decoding methodology and its translation from offline to online operation. Note that the threshold values also affected the performance of simulated random walk (Table~\ref{tab:OnlinePerformances}). In instances where the values of $T_{\mathcal{W}}$ were high, such as in Subjects~A1 and~S1, the random walk simulator had difficulty moving the avatar and consequently could not finish the task within the 20-min time limit. On the other hand, the low values of $T_{\mathcal{I}}$ (e.g. Subjects~A1, ~A2, ~A3 and~A8) resulted in the random walk simulator having difficulty stopping the avatar and therefore yielded low successful stop scores. Finally, when the two thresholds were close to each other and around the chance level, such as in Subjects~A4, ~A5 and~A6, the random walk simulator ``inched'' the avatar forward, thereby achieving high successful stop scores at the expense of longer completion times. In summary, these observations are consistent with the ideal conditions where  $T_{\mathcal{W}}$ and $T_{\mathcal{I}}$ approach the values of 1 and 0, respectively.
They also underscore a trade-off between the completion times and successful stop scores inherent in the design of the online task.

\subsection{Online Performance}   
As shown in Table~\ref{tab:OnlinePerformances}, the subjects' average online performance measures ranged from 228.8 to 410.6 sec for completion time and from 7.65 to 9.34 for number of successful stops, with Subjects~A8 and~S1 achieving the highest number of successful stops. While the performance of Subject~S1 is encouraging, it is unclear whether this generalises to a population of SCI individuals.    
Furthermore, in all but one online session (out of 45), all subjects demonstrated purposeful BCI control. 

While there is a positive correlation between the offline classification accuracy and online performance measures (see Section~\ref{sec:online}), only 21\% of the offline classification variance can be accounted for by the completion times and number of successful stops. This indicates that offline and online performances are only moderately coupled, which may have several underlying causes. First, the high variability of online performances (see Fig.~\ref{fig:MC_results} and Table~\ref{tab:OnlinePerformances}) may cause a poor linear regression fit. Second, a linear regression may not be the best model to link offline and online performances. Finally, the presence of outliers may cause the parameters of the linear regression model to be chosen suboptimally. 
 
As an example of the above discrepancy, the best subject, A8, had an offline performance of only 65\% and yet was able to achieve the level of online control that nearly matched that of a manual joystick. 
This discrepancy 
may be caused by physiological and behavioural factors. 
First, it may be hypothesised that a relatively low offline performance reflects the subject's inconsistency in generating KMI and/or occasional lapse in attention. Since offline training is done without feedback, the subject may not be aware of these issues. Ultimately, this may lead to a decoding model that is suboptimal and hence yields a low offline performance. When online, the feedback is always present, allowing the subject to hone their mental strategy and presumably utilise KMI that is most consistent with the model. The subject's ability to adapt and achieve good performance during online BCI operation may also indicate that the decoding model retains useful KMI features despite being suboptimal.

In general, the BCI performances were inferior to those of a manually-controlled physical joystick.
Note, however, that Subject~A8, a na\"{i}ve BCI user, was on average only 23 sec slower than the joystick with a statistically equivalent number of successful stops. 
Therefore, additional training and online practice may help further reduce the completion time, possibly to the point of approaching that of manual control.
Should this goal be achieved, it could further justify the pursuit of BCI-controlled lower extremity prostheses, whose performances would approach those of manually-controlled prostheses while emulating able-bodied like control. 

\subsection{Conclusions}
In summary, the BCI-controlled walking simulator presented in this study satisfied the three proposed criteria of a practical BCI system, namely intuitiveness, robustness, and short training time. First, the operation of the system was intuitive as it enabled subjects to use walking KMI to control the ambulation of the avatar. Secondly, the system was robust in that the data-driven decoding methodology was able to successfully accommodate for subject-to-subject variations in the neurophysiological underpinnings of idling and walking KMI behaviours (e.g. differences between able-bodied and SCI subjects). Finally, the system required only a short training time, as BCI control was attained after only a 10-min long training data collection procedure followed by a 2-min calibration session. To determine the suitability of the system as a basis for future BCI-controlled lower extremity prostheses to restore ambulation, its function must be further tested in a population of individuals with paraplegia due to SCI. Based on the results achieved in this study, the success of the system in a population of SCI subjects is a realistic proposition. 

\ack
This study was funded by the Roman Reed Spinal Cord Injury Research Fund of California (RR 08-258 and RR 10-281).

\section*{References}
\bibliographystyle{unsrt}
\bibliography{../IEEEabrv,../BCIGaitAvatar}
\end{document}